\documentclass[11pt]{article}
\pdfoutput=1 
\usepackage{amssymb}
\usepackage{amsmath}
\usepackage{amstext}
\usepackage{graphicx,epsfig}
\usepackage{epsfig}
\usepackage{verbatim} 
\usepackage{fancybox}
\usepackage{color}
\usepackage{ulem}
\usepackage{enumitem}
\usepackage{subfigure}
\usepackage{bbm}
\usepackage{parskip}
\usepackage[numbers,sort&compress]{natbib}
\usepackage{ytableau}

\usepackage{tikz}
\usetikzlibrary{decorations}
\pgfdeclaredecoration{complete sines}{initial}
{
    \state{initial}[
        width=+0pt,
        next state=upsine,
        persistent precomputation={\pgfmathsetmacro\matchinglength{
            \pgfdecoratedinputsegmentlength / int(\pgfdecoratedinputsegmentlength/\pgfdecorationsegmentlength)}
            \setlength{\pgfdecorationsegmentlength}{\matchinglength pt}
        }] {}
    \state{upsine}[width=\pgfdecorationsegmentlength,next state=downsine]{
        \pgfpathsine{\pgfpoint{0.25\pgfdecorationsegmentlength}{0.5\pgfdecorationsegmentamplitude}}
        \pgfpathcosine{\pgfpoint{0.25\pgfdecorationsegmentlength}{-0.5\pgfdecorationsegmentamplitude}}
    }
    \state{downsine}[width=\pgfdecorationsegmentlength,next state=upsine]{
        \pgfpathsine{\pgfpoint{0.25\pgfdecorationsegmentlength}{-0.5\pgfdecorationsegmentamplitude}}
        \pgfpathcosine{\pgfpoint{0.25\pgfdecorationsegmentlength}{0.5\pgfdecorationsegmentamplitude}}
}
    \state{final}{}
}

\newcommand{\Comment}[1]{{}}
\definecolor{darkblue}{rgb}{0.15,0.35,0.55}
\definecolor{reddish}{rgb}{0.65, 0.2, 0.2}
\usepackage[linktocpage=true]{hyperref}
\hypersetup{
colorlinks=true,
citecolor=darkblue,
linkcolor=reddish,
urlcolor=darkblue,
pdfauthor={},
pdftitle={},
pdfsubject={}
}

\newcommand{\be}{\begin{equation}}
\newcommand{\ee}{\end{equation}}
\newcommand{\bea}{\begin{eqnarray}}
\newcommand{\eea}{\end{eqnarray}}
\newcommand{\beas}{\begin{eqnarray*}}
\newcommand{\eeas}{\end{eqnarray*}}

\def\({\left(}
\def\){\right)}

\newcommand{\rd}{{\rm d}}

\def\gsim{ \lower .75ex \hbox{$\sim$} \llap{\raise .27ex \hbox{$>$}} }
\def\lsim{ \lower .75ex \hbox{$\sim$} \llap{\raise .27ex \hbox{$<$}} }

\def\xyma{\xymatrix@M.7em}
\def\xymas{\xymatrix@M.1em}
\newcommand{\ba}{\begin{eqnarray}}
\newcommand{\ea}{\end{eqnarray}}

\title{}
\author{}

\begin{document}
%
\renewcommand{\thefootnote}{\fnsymbol{footnote}}
~
\vspace{1.75truecm}
\begin{center}
{\LARGE \bf{Eikonal Scattering and Asymptotic Superluminality}}
\vspace{.25cm}
{\LARGE \bf{of Massless Higher Spin Fields}}
\end{center} 

\vspace{1truecm}
\thispagestyle{empty}
\centerline{{\Large Kurt Hinterbichler,${}^{\rm a,}$\footnote{\href{mailto:kurt.hinterbichler@case.edu}{\texttt{kurt.hinterbichler@case.edu}}} Austin Joyce,${}^{\rm b,}$\footnote{\href{mailto:austin.joyce@columbia.edu}{\texttt{austin.joyce@columbia.edu}}} and Rachel A. Rosen${}^{\rm b,}$\footnote{\href{mailto:rar2172@columbia.edu}{\texttt{rar2172@columbia.edu}}}}}
\vspace{.5cm}
 
\centerline{{\it ${}^{\rm a}$CERCA, Department of Physics,}}
 \centerline{{\it Case Western Reserve University, 10900 Euclid Ave, Cleveland, OH 44106}} 
 \vspace{.25cm}
 
 \centerline{{\it ${}^{\rm b}$Center for Theoretical Physics, Department of Physics,}}
 \centerline{{\it Columbia University, New York, NY 10027}} 
 \vspace{.25cm}

 \vspace{.8cm}
\begin{abstract}
\noindent
We consider scattering of massless higher-spin particles in the eikonal regime in four dimensions.  By demanding the absence of asymptotic superluminality, corresponding to positivity of the eikonal phase, we place constraints on the possible cubic couplings which can appear in the theory.  The cubic couplings come in two types: lower-derivative non-abelian vertices, and higher-derivative abelian vertices made out of gauge-invariant curvature tensors.   We find that the abelian couplings between massless higher spins  lead to an asymptotic time advance for certain choices of polarizations, indicating that these couplings should be absent unless new states come in at the scale suppressing the derivatives in these couplings.  A subset of non-abelian cubic couplings are consistent with eikonal positivity, but are ruled out by consistency of the four-particle amplitude away from the eikonal limit.  The eikonal constraints are therefore complementary to the four-particle test, ruling out even trivial cubic curvature couplings in any theory with a finite number of massless higher spins and no new physics at the scale suppressing derivatives in these vertices.

\end{abstract}

\newpage

\renewcommand*{\thefootnote}{\arabic{footnote}}
\setcounter{footnote}{0}

\section{Introduction}
Interactions of massless higher-spin particles in four dimensional flat spacetime are strongly constrained by classic no-go theorems \cite{Bekaert:2010hw}. Perhaps most prominent amongst these is 
Weinberg's soft theorem~\cite{Weinberg:1965nx}. When combined with the Weinberg--Witten theorem~\cite{Weinberg:1980kq}, as suitably strengthened by \cite{Porrati:2008rm}, these arguments essentially rule out the possibility of massless particles with spin $J\geq 2$ interacting either with a graviton described by Einstein gravity, or with anything that 
interacts with such a graviton~\cite{Porrati:2012rd}. There are also on-shell versions of these arguments which arrive at the same conclusion by 
demanding consistent factorization of four-particle amplitudes into their constituent three-point amplitudes~\cite{Benincasa:2007xk,Schuster:2008nh,Benincasa:2011pg,McGady:2013sga,Elvang:2016qvq}.   This rules out higher-spin theories of the Yang--Mills or Einstein type: those with minimal-derivative cubic vertices which require a nonlinear deformation of the gauge symmetry of the linear theory.  In light of these restrictions, massless high-spin particles won't be discovered interacting with the degrees of freedom that we observe in nature, unless our understanding of quantum field theory is deeply flawed. 

Nevertheless, it is important to understand not only the quantum field theories realized in nature but also the space of possible consistent theories. 
The aforementioned four-particle test renders inconsistent a wide swath of higher-spin theories, but there remains a class of higher-spin theories that are unconstrained by all of these no-go arguments: massless higher-spin particles interacting through their linearized curvature tensors.
 A massless spin $J$ field has a $J$-derivative linear curvature tensor which is invariant under the linearized higher spin's gauge symmetries~\cite{deWit:1979sib}.  Interactions can trivially be constructed using products of these linearized curvatures.  These are all higher-derivative interactions.  They do not include anything that might be called minimal coupling to gravity and they do not survive the soft limit, {\it i.e.}, they cannot mediate long range forces.  This is all in agreement with the Weinberg soft theorem and the other no-go theorems.

These curvature interactions are unconstrained by all of the gauge invariance requirements and 4-particle factorization constraints that rule out the non-abelian vertices.
It is therefore of interest to develop different complementary constraints to constrain these curvature interactions.  The constraint we will use 
 comes from positivity of eikonal scattering amplitudes.  Like the 4-particle test, this is an on-shell $S$-matrix criterion.  As such it is an unambiguous 
constraint, independent of the Lagrangian, field re-definitions, {\it et cetera}. It is typically imposed because of its relation to superluminality in shockwave backgrounds~\cite{tHooft:1987vrq,Kabat:1992tb} and leads to nontrivial constraints on higher-curvature corrections to Einstein gravity~\cite{Camanho:2014apa} and on theories of massive spin-2 particles~\cite{Camanho:2016opx,Hinterbichler:2017qyt,Bonifacio:2017nnt}. Though we do not have a fundamenal derivation of eikonal positivity from, {\it e.g.}, analyticity or microlocality in flat space, it is satisfied in known ultraviolet-complete examples~\cite{DAppollonio:2015fly}.\footnote{In anti-de Sitter space it is related to positivity constraints on dual operator anomalous dimensions, which follow from bootstrap-type considerations~\cite{Afkhami-Jeddi:2016ntf,Li:2017lmh,Kulaxizi:2017ixa,Afkhami-Jeddi:2017rmx}.}  Therefore, even in the absence of a fundamental derivation, it is worthwhile to determine the constraints that positivity of the eikonal phase places on theories of massless higher-spin particles.

Here we investigate eikonal scattering of massless higher-spin particles in four dimensions. In the cubic vertices we introduce two mass scales: $M$ and $\Lambda$.  The scale $M$ suppresses powers of the field  and sets the scale of loop corrections while $\Lambda$ is the scale suppressing derivatives.  Weak coupling requires $\Lambda\ll M$.  The lowest scale suppressing interactions in this theory is denoted by \linebreak $\Lambda_c = (M \Lambda^{J_1+J_2+J_3-2})^\frac{1}{J_1+J_2+J_3-1}$ where $J_n$ represents the spin of each of the three particles.  This is where we would expect tree level unitarity to break down and thus $\Lambda_c$ sets the cutoff of the effective field theory (EFT).  In what follows we show that higher-spin theories interacting at cubic order through their linearized gauge-invariant curvatures experience asymptotic time advances for certain choices of polarizations.   The assumption that consistent theories should not have measurable time advances thus renders these cubic vertices inconsistent, unless new physics enters at the scale $\Lambda$ which is generically much lower than $\Lambda_c$.  In particular, any interacting theory containing only a finite number of massless higher spins cannot have any cubic vertices if no new physics enters at $\Lambda$ (though it may be possible with an infinite number of massless higher spins).

Interestingly, these constraints are complementary to those coming from the four-particle test. Eikonal constraints restrict couplings through curvatures, but only partially constrain the minimal-derivative couplings which require a nonlinear deformation of the linear gauge symmetry. This work generalizes the case of the massless spin-2 field: there, the minimal derivative vertex is constrained by the 4-particle test to be that of Einstein-Hilbert and to satisfy the equivalence principle, and never gives a time advance (rather, it gives the Shapiro delay), whereas the Riemann cubed vertex is unconstrained by the 4-particle test but is constrained to vanish by asymptotic superluminality, unless new states come in~\cite{Camanho:2014apa}.\footnote{See also \cite{Henneaux:2013vca,Hertzberg:2016djj,Hertzberg:2017abn} for similar arguments and conclusions}

Taken together, our constraints and those of the classic no-go theorems rule out all interacting massless higher-spin vertices in four dimensions. This closes a loophole in various arguments against higher-spin particles, which either explicitly or implicitly assume that higher spins interact through vertices which require a nonlinear completion of the gauge symmetry.

\section{Eikonal scattering}
The eikonal regime of $2\to2$ scattering is the kinematic regime where the center of mass energy is taken to be large with the impact parameter held fixed.  The eikonal kinematics for massless particles are reviewed in Appendix~\ref{app:polarizations}.  With these kinematics, the leading contribution to scattering is given by the sum of ladder and crossed ladder diagrams in the $t$-channel. These diagrams exponentiate in impact parameter space and the amplitude in the eikonal limit is given by (suppressing polarization labels)~\cite{Cheng:1969eh,Levy:1969cr,Abarbanel:1969ek}
\be
i{\cal M}_{\rm eikonal}(s, t) = 2s\int\rd^{2}\vec b\,e^{i\vec q\cdot\vec b}\left(e^{i\delta(s,\vec b)}-1\right),
\ee
where $\vec b$ is the impact parameter, the Fourier conjugate to the exchanged momentum $\vec q$, and
where the eikonal phase, $\delta$, is given by
\be
\delta(s,\vec b) = \frac{1}{2s}\int\frac{\rd^{2}\vec q}{(2\pi)^{2}}\,e^{-i\vec q\cdot\vec b}{\cal M}_{4}(s, -\vec q\,{}^2)\,,
\label{eq:eikphaseint}
\ee
with ${\cal M}_{4}(s,t)$ the tree-level $t$-channel amplitude evaluated in the eikonal limit.
The eikonal approximation captures the leading interactions between two highly boosted particles.  The eikonal phase is related to the delay in lightcone coordinate time, $\Delta x^{\scriptscriptstyle -}$, experienced by the particle moving in the $x^{\scriptscriptstyle +}$ direction after interacting with the other particle moving in the $x^{\scriptscriptstyle -}$ direction~\cite{Kabat:1992tb,Camanho:2014apa,Hinterbichler:2017qyt}
\be
\Delta x^{\scriptscriptstyle -} = \frac{1}{\lvert p^{\scriptscriptstyle -}\rvert}\delta(s, b).
\ee
The expectation is that, in theories which obey microcausality, particles will only ever experience time delays ($\Delta x^{\scriptscriptstyle -}\geq 0$) from interactions, corresponding to positivity of the eikonal phase, $\delta >0$. We therefore want to see what constraints positivity of the eikonal phase places on theories containing massless higher-spin particles.

The leading eikonal amplitude only depends on the on-shell three-point scattering amplitudes, which can be seen by performing a complex deformation of the integration contour in~\eqref{eq:eikphaseint}~\cite{Camanho:2014apa}.  This contour integral picks up a $t$-channel pole, and on this complex factorization channel the exchanged particle goes on-shell and the eikonal phase is given by a product of on-shell three point amplitudes, which can be represented as differential operators acting on the propagator of the exchanged particle.  This results in the expression
\be
\delta(s,b) = \frac{\sum_I{\cal M}_3^{13 I}(i\partial_{\vec b}){\cal M}_3^{I24}(i\partial_{\vec b})}{2s}\int\frac{\rd^{2}\vec q}{(2\pi)^{2}}\frac{e^{-i\vec q\cdot\vec b}}{\vec q^2} = \frac{\sum_I{\cal M}_3^{13 I}(i\partial_{\vec b}){\cal M}_3^{I24}(i\partial_{\vec b})}{2s}\frac{1}{2\pi}\log\left(\frac{L}{b}\right), \label{eikonalampde}
\ee
where the $I$ index sums over the polarizations of the exchanged particle. In the last equality we have introduced an infrared regulator, $L$.\footnote{The presence of a logarithmic dependence in the eikonal phase might be somewhat distressing. At large impact parameter, $b > L$, this appears to lead to a negative phase shift. This is related to to infrared divergences inherent in massless scattering in four dimensions. More properly, the eikonal phase should be compared to the same phase in, {\it e.g.}, Einstein gravity and should be positive compared to this. This is effectively the criterion of~\cite{Gao:2000ga} that signals cannot propagate faster than allowed by the asymptotic causal structure of a spacetime. In practice we will be concerned with the small impact parameter behavior of the eikonal phase, where these infrared effects are irrelevant.}

\section{Cubic couplings of massless higher spins}
 We want to compute the eikonal amplitude \eqref{eikonalampde}, which requires enumerating the possible cubic vertices for massless higher spins.
The cubic couplings between higher-spin particles are so strongly constrained by Lorentz invariance that there are only a finite number of possible couplings between particles with spins $J_1, J_2, J_3$. These three-point amplitudes can be constructed using many different approaches~\cite{Bengtsson:1983pd,Berends:1984wp,Berends:1984rq,
 Bengtsson:1986kh,
 Metsaev:1991mt,Metsaev:1991nb,Metsaev:1993ap,Metsaev:2005ar,Benincasa:2007xk,
 Fotopoulos:2008ka,Zinoviev:2008ck,Boulanger:2008tg,Manvelyan:2010jr,Sagnotti:2010at,Fotopoulos:2010ay,
 Manvelyan:2010je,Costa:2011mg,Conde:2016izb,Francia:2016weg}. The degrees of freedom of the $a$-th massless particle ($a=1,2,3$) are carried by a transverse-traceless polarization tensor $\epsilon_{\mu_1\cdots \mu_{J_a}}$.  It is convenient to introduce auxiliary variables, $z_a^\mu$, which are null ($z_a^2 = 0$) and transverse ($p_a\cdot z_a = 0$).  After constructing the on-shell vertices in terms of the $z$'s, we can make the replacement $z_a^{\mu_1}\cdots z_a^{\mu_{J_a}}\mapsto \epsilon_a^{\mu_1\cdots \mu_{J_a}}$ to reintroduce the polarization tensors. The scattering amplitude with spins $\{J_1,J_2,J_3\}$ must be homogeneous of order $J_a$ in each of the $z_a$.  Gauge invariance in this language corresponds to invariance under the shift 
$
z_a\mapsto z_a + \epsilon\, p_a.
$
After accounting for momentum conservation ($\sum_a p^\mu_a=0$), and the masslessness of external particles ($p_a^2=0$), a convenient set of building blocks for massless on-shell cubic amplitudes is given by (all momenta ingoing)
\begin{align}
\label{eq:astruc}
{\cal A} &=(p_1\cdot z_3)(z_1\cdot z_2)+(p_3\cdot z_2)(z_1\cdot z_3)+(p_2\cdot z_1)(z_2\cdot z_3)\, ,\\
\label{eq:bstruc}
{\cal B} &= (p_1\cdot z_3)(p_2\cdot z_1)(p_3\cdot z_2)\, ,\\
{\cal C}_a &= (p_{a+1}\cdot z_a)\, .
\end{align}
The structures ${\cal A}$ and ${\cal B}$ are anti-symmetric in the $1,2,3$ labels (they are just the two independent cubic vertices of three massless spin-1 particles) while ${\cal C}_a$ is anti-symmetric in the arguments $\neq a$ (it is the spin-0--spin-0--spin-1 cubic vertex).

If we order the three spins as (without loss of generality) $J_1\leq J_2\leq J_3$, the list of possible gauge-invariant three-point structures is
\begin{align}
\nonumber
{\cal M}_1 &= {\cal A}^{J_1} {\cal C}_2^{J_2-J_1}{\cal C}_3^{J_3-J_1}\, ,\\\nonumber
{\cal M}_2 &= {\cal A}^{J_1-1}{\cal B}\,{\cal C}_2^{J_2-J_1}{\cal C}_3^{J_3-J_1}\, , \\\nonumber
&~\,\vdots\\\nonumber
{\cal M}_{j+1} &= {\cal A}^{J_1-j}{\cal B}^j\,{\cal C}_2^{J_2-J_1}{\cal C}_3^{J_3-J_1}\, ,\\\nonumber
&~\,\vdots\\
{\cal M}_{J_1+1} &= {\cal B}^{J_1}{\cal C}_2^{J_2-J_1}{\cal C}_3^{J_3-J_1} ={\cal C}_1^{J_1}{\cal C}_2^{J_2}{\cal C}_3^{J_3}\, .
\label{eq:curvcubedcoupling}
\end{align}
There are $J_1+1$ possible vertices, which start with $J_2+J_3-J_1$ derivatives and count up by two up to a vertex with $J_1+J_2+J_3$ derivatives.

A massless spin-$J$ field possesses a linearized curvature tensor of the form (see, {\it e.g.},~\cite{Sorokin:2004ie})
 \be
{\cal F}_{\mu_1\nu_2\cdots\mu_J\nu_J} \sim {\cal Y}_{[J,J]} \partial_{\mu_1}\partial_{\mu_2}\cdots\partial_{\mu_J}\ell_{\nu_1\cdots\nu_J},
\ee
where ${\cal Y}_{[J,J]}$ is the Young projector onto the Young diagram~\scalebox{.45}{$
\begin{array}{|c c c c c|}\hline
&&J&&\\
\hline
&&J&&\\
\cline{1-5}
\end{array}
$}\,.
This curvature is invariant under the linearized gauge transformation of the spin-$J$ field,
\be
\delta \ell_{\mu_1\cdots\mu_J} = \partial_{(\mu_1}\Lambda_{\mu_2\cdots\mu_J)}.
\ee
Here $\ell_{\mu_1\cdots\mu_J}$ is the symmetric double-traceless spin-$J$ field in the Fronsdal formulation \cite{Fronsdal:1978rb}, and 
$\Lambda_{\mu_1\cdots\mu_{J-1}}$ its symmetric traceless gauge parameter. The maximal-derivative coupling, ${\cal M}_{J_1+1}$, is the one resulting from the product of three of these generalized curvatures. It is these couplings which are unconstrained by four-particle consistency but which will be constrained by eikonal scattering.

In four dimensions, not all of the 3-point amplitudes in \eqref{eq:curvcubedcoupling} are nontrivial. This is because the amplitudes depend on the 5 vectors $\{ p_1, p_2, z_1, z_2, z_3\}$, but in four dimensions, it is not possible for 5 vectors to be linearly independent. There is an identity which expresses this fact~\cite{Conde:2016izb}
\be
\epsilon_{\mu_1\mu_2\mu_3\mu_4\mu_5}\epsilon^{\nu_1\nu_2\nu_3\nu_4\nu_5} p_1^{\mu_1}p_{1 \,\nu_1}p_2^{\mu_2}p_{2\,\nu_2}z_1^{\mu_3}z_{1\,\nu_3}z_2^{\mu_4}z_{2\,\nu_4}z_3^{\mu_5}z_{3\,\nu_5} = 0.
\ee
After using the on-shell conditions and the fact that $z$ is transverse and null, this identity reduces to the statement that ${\cal A}{\cal B} = 0$,
where ${\cal A}$ and ${\cal B}$ are the structures defined in~\eqref{eq:astruc} and~\eqref{eq:bstruc}, respectively.  This means that in four dimensions all the middle structures in the list \eqref{eq:curvcubedcoupling} identically vanish and we are left with only ${\cal M}_1$ and ${\cal M}_{J_1+1}$.
We can think of of ${\cal M}_1$ as the analogue of the Einstein--Hilbert vertex, which deforms the gauge symmetry and has $J_2+J_3-J_1$ derivatives, whereas ${\cal M}_{J_1+1}$ is the curvature-cubed interaction vertex which has $J_1+J_2+J_3$ derivatives.  The middle vertices are the analogues of the Gauss--Bonnet vertex which is trivial in four dimensions.  In what follows we will specialize to the four dimensional case.   In this case there are also some parity-odd vertices which become non-trivial; we will impose parity and will not consider these.

\section{Eikonal amplitude}
Let's consider the general situation where three particles of spins $J_1\leq J_2\leq J_3$ interact through an arbitrary linear combination of the two vertices ${\cal M}_1$ and ${\cal M}_{J_1+1}$. These vertices are
\begin{align}
\label{eq:d4vert2}
\nonumber
{\cal M}_1 &= \frac{1}{M \Lambda^{J_2+J_3-J_1-2}} \\
& \times \left[(p_1\cdot z_3)(z_1\cdot z_2)+(p_3\cdot z_2)(z_1\cdot z_3)+(p_2\cdot z_1)(z_2\cdot z_3)\right]^{J_1}(p_{3}\cdot z_2)^{J_2-J_1}(p_{1}\cdot z_3)^{J_3-J_1}\, ,\\
{\cal M}_{J_1+1} &= \frac{1}{M\Lambda^{J_1+J_2+J_3-2}}(p_{2}\cdot z_1)^{J_1}(p_{3}\cdot z_2)^{J_2}(p_{1}\cdot z_3)^{J_3}\, . \nonumber
\end{align}

We have introduced two mass scales, $\Lambda$ and $M$.  The scale $M$ is the analogue of a Planck mass; it suppresses powers of the field relative to the kinetic term and sets the scale of loop corrections.\footnote{It is worth emphasizing, however, that we are {\it not} making any assumptions about the power-counting structure of the theory beyond cubic order. Even in a situation where derivatives in other interactions are suppressed by a scale parametrically above $\Lambda$, our analysis applies and new physics must enter at the scale $\Lambda$ if a time advance is possible.}  $\Lambda$ is a scale suppressing derivatives; if the higher derivative term were the result of integrating out massive particles, $\Lambda$ would be the mass of these particles.  As discussed in \cite{Camanho:2014apa}, the regime in which we work is the regime in which energies are large compared to $\Lambda$ so that we are in the eikonal regime and probing the non-linearities of the higher derivative terms, but smaller than $M$ (or any intermediate strong coupling scale) so that we remain weakly coupled.  Thus we must have $\Lambda\ll M$.  If we find that a vertex leads to a negative phase shift, it means that new physics must come in at the scale $\Lambda$ in order to cure it, otherwise the vertex must be zero.\footnote{More precisely, its coefficient must be sufficiently small as to make it effectively suppressed by the scale $M$ rather than $\Lambda$.}

The general 4-point interaction constructed from these vertices is somewhat complicated, because any of the three constituent particles can appear on the internal line. However, in the eikonal regime the contribution from spin-$J$ exchange grows as $\delta \sim s^{J-1}$ \cite{Camanho:2014apa}, and so the amplitude is dominated by exchange of the particle with the largest spin, $J_3$.  

The leading eikonal phase then arises from the $t$-channel exchange of the spin-$J_3$ particle:
\vspace{-.7cm}
\be
\begin{tikzpicture}[line width=1.7 pt,baseline={([yshift=-3ex]current bounding box.center)},vertex/.style={anchor=base,
    circle,fill=black!25,minimum size=18pt,inner sep=2pt}]
\draw[style={decorate,decoration=complete sines},line width=1] (-1,0) -- (3,0);
\draw[style={decorate,decoration=complete sines},line width=1] (-1,-.08) -- (3,-.08);
\draw[style={decorate,decoration=complete sines},line width=1] (-1,1.06) -- (3,1.06);
\draw[style={decorate,decoration=complete sines},line width=1] (-1,1.14) -- (3,1.14);
\draw[style={decorate,decoration=complete sines},line width=1] (-.04,0) -- (-.04,2);
\draw[style={decorate,decoration=complete sines},line width=1] (.04,0) -- (.04,2);
\node[scale=1] at (1.9, .15) {$,$};
\node[scale=1] at (1.4, -.03) {$J_2,4$};
\node[scale=1] at (-1.4, -.03) {$2,J_1$};
\node[scale=1] at (-1.4, 1.15) {$1,J_1$};
\node[scale=1] at (1.4, 1.1) {$J_2,3$};
\node[scale=1] at (-3,0) {$~$};
\node[scale=1] at (.3, .5) {$J_3$};
\end{tikzpicture}
\ee
along with three other diagrams which interchange the spin $J_1$ and $J_2$ particles on either the top rail, the bottom rail, or both.\footnote{Under interchanging the $1$ and $2$ particles, the amplitudes in~\eqref{eq:d4vert2} pick up a factor of $(-1)^{J_3}$. In the case that this is negative, there must be appropriate color factors in order to restore the correct bosonic statistics. In what follows we leave these factors implicit, but ensure that the resulting eikonal amplitude is appropriately symmetric.}
In the eikonal limit the lightcone momenta are taken to be large, $p^+ ,p^-\to\infty$. A drastic simplification in this limit follows from the fact that the momentum on the external lines is much larger than the internal rung momentum. This means that the the $N_{+_1\cdots +_{J_3} -_1\cdots -_{J_3}}$ component of the internal propagator~\eqref{eq:spinJprop} gives the leading contribution. In practice this means that $z_I^\mu$ points along the $+$ direction in the top vertex and in the $-$ direction in the bottom vertex. Putting the vertices on-shell and taking the eikonal limit, we obtain the product of on-shell cubic amplitudes from the sum of all $t$-channel diagrams 
\begin{align}
\nonumber
\sum_I{\cal M}_3^{13I}{\cal M}_3^{I24}  =&\, \frac{s^{J_3}}{2^{J_3}M^2} \Big[ \frac{1}{ \Lambda^{J_2+J_3-J_1-2}}\left(e_1^{i_1\cdots i_{J_1}}e_3^{i_1\cdots i_{J_1} j_{J_1+1}\cdots j_{J_2}} +e_3^{i_1\cdots i_{J_1}}e_1^{i_1\cdots i_{J_1} j_{J_1+1}\cdots j_{J_2}}\right) q_{j_{J_1+1}}\cdots q_{j_{J_2}}\\\nonumber
&~~~~~~+\frac{g(-1)^{J_1}}{ \Lambda^{J_1+J_2+J_3-2}}\left( e_1^{i_1\cdots i_{J_1}}e_3^{j_1\cdots j_{J_2}}+e_3^{i_1\cdots i_{J_1}}e_1^{j_1\cdots j_{J_2}}\right)q_{i_1}\cdots q_{i_{J_1}}q_{j_1}\cdots q_{j_{J_2}}\Big]\\\nonumber
&\times\Big[ \frac{(-1)^{J_2-J_1}}{ \Lambda^{J_2+J_3-J_1-2}}\left(e_2^{i_1\cdots i_{J_1}}e_4^{i_1\cdots i_{J_1} j_{J_1+1}\cdots j_{J_2}} +e_4^{i_1\cdots i_{J_1}}e_2^{i_1\cdots i_{J_1} j_{J_1+1}\cdots j_{J_2}} \right)q_{j_{J_1+1}}\cdots q_{j_{J_2}} \\
&~~~~~~+\frac{g(-1)^{J_2}}{ \Lambda^{J_1+J_2+J_3-2}}\left( e_2^{i_1\cdots i_{J_1}}e_4^{j_1\cdots j_{J_2}}+ e_4^{i_1\cdots i_{J_1}}e_2^{j_1\cdots j_{J_2}}\right)q_{i_1}\cdots q_{i_{J_1}}q_{j_1}\cdots q_{j_{J_2}}\Big].
\end{align}
Here $g$ is a (dimensionless) coupling constant to distinguish the ${\cal M}_{J+1}$ vertex and we have normalized the ${\cal M}_1$ vertex to have unit coefficient.
From this expression, It is straightforward to write down the phase shift using \eqref{eikonalampde},
\begin{align}
\nonumber
\delta(s,b) =&\, \frac{s^{J_3-1}}{2^{J_3+1}M^2} \Big[ \frac{1}{ \Lambda^{J_2+J_3-J_1-2}}\left(e_1^{i_1\cdots i_{J_1}}e_3^{i_1\cdots i_{J_1} j_{J_1+1}\cdots j_{J_2}} +e_3^{i_1\cdots i_{J_1}}e_1^{i_1\cdots i_{J_1} j_{J_1+1}\cdots j_{J_2}}\right) \partial_{b_{j_{J_1+1}}}\cdots \partial_{b_{j_{J_2}}}\\
&~~~~~~+\frac{g}{ \Lambda^{J_1+J_2+J_3-2}}\left( e_1^{i_1\cdots i_{J_1}}e_3^{j_1\cdots j_{J_2}}+e_3^{i_1\cdots i_{J_1}}e_1^{j_1\cdots j_{J_2}}\right)\partial_{b_{i_1}}\cdots \partial_{b_{i_{J_1}}}\partial_{b_{j_1}}\cdots \partial_{b_{j_{J_2}}}\Big]\\\nonumber
&\times\Big[ \frac{1}{ \Lambda^{J_2+J_3-J_1-2}}\left(e_2^{i_1\cdots i_{J_1}}e_4^{i_1\cdots i_{J_1} j_{J_1+1}\cdots j_{J_2}} +e_4^{i_1\cdots i_{J_1}}e_2^{i_1\cdots i_{J_1} j_{J_1+1}\cdots j_{J_2}} \right)\partial_{b_{j_{J_1+1}}}\cdots \partial_{b_{j_{J_2}}} \\\nonumber
&~~~~~~+\frac{g}{ \Lambda^{J_1+J_2+J_3-2}}\left( e_2^{i_1\cdots i_{J_1}}e_4^{j_1\cdots j_{J_2}}+ e_4^{i_1\cdots i_{J_1}}e_2^{j_1\cdots j_{J_2}}\right)\partial_{b_{i_1}}\cdots \partial_{b_{i_{J_1}}}\partial_{b_{j_1}}\cdots \partial_{b_{j_{J_2}}}\Big]\frac{1}{2\pi}\log \left(\frac{L}{b}\right)\,.
\end{align}
In order to ascertain whether a particle will experience a positive or negative phase shift in scattering, we should diagonalize $\delta$ to determine the phase shifts of the eigen-polarizations.
At sufficiently small impact parameter, the amplitude is dominated by:
\begin{align}
\nonumber
\delta \simeq g^2\frac{s^{J_3-1}}{2^{J_3+1}M^2\Lambda^{2J_1+2J_2+2J_3-4}}&
\left( e_1^{i_1\cdots i_{J_1}}e_3^{j_1\cdots j_{J_2}}+e_3^{i_1\cdots i_{J_1}}e_1^{j_1\cdots j_{J_2}}\right)\left( e_2^{k_1\cdots k_{J_1}}e_4^{l_1\cdots l_{J_2}}+ e_4^{k_1\cdots k_{J_1}}e_2^{l_1\cdots l_{J_2}}\right)\\
&\!\!\!\!\!\!\!\times\partial_{b_{i_1}}\cdots \partial_{b_{i_{J_1}}}\partial_{b_{j_1}}\cdots \partial_{b_{j_{J_2}}}\partial_{b_{k_1}}\cdots \partial_{b_{k_{J_1}}}\partial_{b_{l_1}}\cdots \partial_{b_{l_{J_2}}}
\frac{1}{2\pi}\log \left(\frac{L}{b}\right).
\label{eq:curvcubedslJ}
\end{align}
There are two possible polarization choices for each of the incoming and outgoing particles, and the most general possible scattering process involves an admixture of particles with spin $J_1$ and $J_2$ carrying a linear combination of their two possible polarizations in all the external lines. Utilizing the basis of polarizations described in Appendix~\ref{app:polarizations} we can diagonalize this amplitude, as described in Appendix~\ref{app:eikampcalc}. At the end, we find the eigen-shifts:
\be
\label{eq:leadingeigenvalues}
\delta(s,b) \simeq \pm g^2\frac{(4(J_1+J_2)-2)!!}{ 2^{J_{1}+J_2+J_3+1}\pi}\frac{\Lambda^2}{M^2}\left(s\over \Lambda^2\right)^{J_3-1} \frac{1}{(\Lambda^2 b^2)^{J_1+J_2}},
\ee
with multiplicity 4. We see that regardless of the sign of the coupling, $g$, some polarizations will experience a time advance. Such a time advance will be measurable at impact parameters $b\sim \Lambda^{-1}$, which is parametrically within the regime of validity of the effective theory. Therefore, we must either set the coupling $g$ to be zero or some new physics must enter at the scale $\Lambda$, which is parametrically below the scale at which perturbative unitarity is lost.

After setting $g = 0$, the phase shift is given by
\begin{align}
\label{eq:finaleikamp}
\delta(s,b) &=\, \frac{s^{J_3-1}}{2^{J_3+1}M^2 \Lambda^{2J_2+2J_3-2J_1-4}}\left(e_1^{i_1\cdots i_{J_1}}e_3^{i_1\cdots i_{J_1} j_{J_1+1}\cdots j_{J_2}} +e_3^{i_1\cdots i_{J_1}}e_1^{i_1\cdots i_{J_1} j_{J_1+1}\cdots j_{J_2}}\right)\\\nonumber
&\times\left(e_2^{i_1\cdots i_{J_1}}e_4^{i_1\cdots i_{J_1} j_{J_1+1}\cdots j_{J_2}} +e_4^{i_1\cdots i_{J_1}}e_2^{i_1\cdots i_{J_1} j_{J_1+1}\cdots j_{J_2}} \right)\partial_{b_{j_{J_1+1}}}\cdots \partial_{b_{j_{J_2}}}\partial_{b_{j_{J_1+1}}}\cdots \partial_{b_{j_{J_2}}}\frac{1}{2\pi}\log \left(\frac{L}{b}\right)\,.
\end{align}
In the case $J_2 = J_1$, the amplitude is diagonal, with the phase shifts 
\be
\delta_{J_2 = J_1} (s,b)= \frac{1}{2^{J_3+2}\pi}\frac{\Lambda^2}{M^2}\left(\frac{s}{\Lambda^2}\right)^{J_3-1}\log \left(\frac{L}{b}\right),
\label{eq:equalsubleading}
\ee
which are positive, consistent with asymptotic causality. In the more general case where $J_2 > J_1$, a further computation is required, which is detailed in Appendix~\ref{app:eikampcalc}, leading to the phase shifts
\be
\delta_{J_2 > J_1} (s,b)= \pm \frac{(2(J_2-J_1)-2)!!}{2^{J_3+J_2-J_1+1}\pi}\frac{\Lambda^2}{M^2}\left(\frac{s}{\Lambda^2}\right)^{J_3-1}\frac{1}{(\Lambda^2 b^2)^{J_2-J_1}}
\label{eq:subleadingphase}
\ee
with multiplicity 8. Here we see that some polarization combinations will experience a time advance, so even non-abelian couplings between massless higher spins must be set to zero if $J_2 > J_1$. 

The net result is that we have to set the curvature-cubed couplings to zero; in the case of non-abelian couplings, we are allowed to have cubic couplings which deform the gauge symmetry between two particles with the same spin, and a particle with larger or equal spin.\footnote{Although there does not exist a fully gauge-invariant $S$-matrix for vertices which deform the gauge symmetry involving spins $J>2$, the eikonal amplitude is only sensitive to the on-shell cubic vertices in the $t$-channel exchange, which do not know about the failure of gauge invariance at quartic order.  Thus the eikonal amplitude using these minimal derivative vertices can be made sense of and is consistent with positivity of the eikonal phase.}  Other non-abelian cubic couplings must also be set to zero.

\section{Conclusions}
We have enumerated the constraints that positivity of the eikonal phase places on theories of interacting massless higher-spin particles in four dimensions. These constraints are fairly strong; they rule out cubic interactions between particles through their linear gauge invariant curvature tensors, unless new physics appears at the scale suppressing their derivatives. Additionally, amongst cubic couplings which require deformations of the linearized gauge symmetries, only interactions between particles of the same spin through exchange of a higher-spin particle are consistent with having an asymptotic time delay. 
These constraints are complementary to other arguments against higher-spin cubic interactions, which rule out interactions that require a deformation of the linearized gauge symmetry. Taken together, this rules out all cubic interactions between massless higher-spin particles in four dimensions. 

There remain some loopholes to this general no-go.  There is still room for curvature couplings between massless higher spins which start at fourth order and are therefore not captured by the leading-order eikonal analysis.  It would be interesting to see if these could be constrained by considering sub-leading corrections to the eikonal approximation.\footnote{It is also possible that a refinement of our argument could rule out some of the non-abelian couplings that experience a time delay, but are known to be inconsistent from the four-particle arguments (for example the coupling of two massless spin-2 particles to a massless spin-4 particle). In particular we have considered the leading high-energy behavior, at smaller values of $s$ it seems possible that these couplings could lead to time advances.}
We have restricted our attention to four dimensions which reduces the number of relevant three-point vertices to two. In higher dimensions, there are more possibilities. To the extent that scattering in higher dimensions can be localized to a four-dimensional subspace, we expect our arguments to hold for the highest-derivative vertices, but it is possible that some of the intermediate vertices are compatible with eikonal positivity.   Finally, it is possible that in a hypothetical flat space higher spin theory with an infinite tower of all higher spins, along the lines considered in \cite{Metsaev:1991mt,Ponomarev:2016jqk,Roiban:2017iqg,Tseytlin:2017xml}, there could be miraculous re-summations among the whole tower that render the eikonal phase positive, and it would be interesting to explore this possibility.

\vspace{-.2cm}
\paragraph{Acknowledgements:}  RAR is supported by DOE grant DE-SC0011941 and Simons Foundation Award Number 555117.  AJ and RAR are supported by NASA grant NNX16AB27G.

\appendix
\section{Eikonal kinematics and polarizations}
\label{app:polarizations}
Here we collect some expressions used in the scattering computation described in the main text.  We work in lightcone coordinates $(x^{\scriptscriptstyle -},x^{\scriptscriptstyle +},x^i)$,
$x^{\scriptscriptstyle \pm}={1\over \sqrt{2}}\left(x^0\pm x^1\right)$, 
where the Minkowski metric takes the form,
\be \eta_{\mu\nu}=\left(\begin{array}{ccc}0 & -1 & 0 \\-1 & 0 & 0 \\0 & 0 & \delta_{ij}\end{array}\right),\ee
with $i,j=1,2$ running over the transverse directions.

The amplitude is computed with the following kinematics which are well-adapted to the eikonal limit, 
\begin{align}
\label{eq:mom1}
p_1^\mu&=\left(p^+,{{\vec q\ }^2\over 8p^+} ,{ q^i \over 2}\right)\, , & p_2^\mu&=\left({{\vec q\ }^2\over 8p^-}, p^-,-{ q^i\over 2 }\right)\, ,\\
p_3^\mu&=\left(p^+,{{\vec q\ }^2\over 8p^+},-{ q^i \over 2}\right)\, ,&p_4^\mu &=\left({{\vec q\ }^2\over 8p^-}, p^-,{ q^i\over 2 }\right)\,.
 \label{eq:mom4}
\end{align}
These square to zero: $p_1^2=p_3^2=p_2^2=p_4^2=0$, and conserve momentum $p_1^\mu+p_2^\mu=p_3^\mu+p_4^\mu$.  The independent Mandelstam invariants are
\bea &&s=-(p_1+p_2)^2=2p^+p^-+{{\vec q\ }^2\over 2}+{1\over 2 p^+ p^-}{{\vec q\ }^4\over 16},\\
&&t=-(p_1-p_3)^2=-{\vec q\ }^2\, .\eea

We construct polarization tensors out of the following transverse spin-1 polarization tensors:
\begin{align}
\label{eq:pol1}
 \epsilon_T^\mu(p_1)&=\left(0,{{\vec q \ }\cdot {\vec e}_1 \over 2p^+} ,{  e_1^i  }\right)\, , &\epsilon_T^\mu(p_2)&=\left(-{{\vec q \ }\cdot  {\vec e}_2 \over 2p^-},0 ,{  e_2 ^i   }\right)\, , \\
\epsilon_T^\mu(p_3)&=\left(0,-{{\vec q \ }\cdot  {\vec e}_3 \over 2p^+} ,{  e_3^i   }\right)\, ,   &\epsilon_T^\mu(p_4)&=\left({{\vec q \ }\cdot  {\vec e}_4 \over 2p^-},0 ,{  e_4^i   }\right)\,  .
 \label{eq:pol4}
\end{align}
Here the $e^i$ are normalized vectors that point in the plane transverse to $x^+,x^-$; there are $2$ independent such vectors. To construct the massless spin-$J$ polarization tensors, we take the product
\be
\epsilon_T^{\mu_1\cdots\mu_J}(p_a) =\epsilon_T^{\mu_1}(p_a)\cdots\epsilon_T^{\mu_J}(p_a)\, ,
\ee
and in this expression we replace
\be
e_{i_1}\cdots e_{i_J}\mapsto e_{i_1\cdots i_J},
\ee
where $e_{i_1\cdots i_J}$ is a symmetric, traceless tensor with indices that take on two values, so it is an SO$(2)$ representation and therefore has two independent helicity components.

The eikonal limit then corresponds to the limit of large $p^+,p^-$.

We also need the spin-$J$ propagator numerator, the only part needed in the eikonal limit is
\be
N_{\mu_1\cdots\mu_J}^{\alpha_1\cdots\alpha_J} = \delta^{(\alpha_1}_{(\mu_1}\cdots\delta_{\mu_J)}^{\alpha_J)} +\cdots.
\label{eq:spinJprop}
\ee
Diagonalizing the eikonal phase requires an explicit basis of helicity-$J$ polarizations. This means we need a basis for traceless symmetric tensors $e_{i_1\cdots i_J}$ in two dimensions. One construction is the following: take as a basis of the plane transverse to $p^+, p^-$ the standard unit vectors
\be
e^{(1)}_i = 
\left(
\begin{array}{c}
1\\
0
\end{array}
\right)\, ,~~~~~~~~~~~~~~
e^{(2)}_i = 
\left(
\begin{array}{c}
0\\
1
\end{array}
\right)\, .
\ee
If we take the linear combinations of these unit vectors
\be
e^{(\pm)}_i  = \frac{1}{\sqrt 2}\left(e^{(1)}_i\pm i e^{(2)}_i\right),
\ee
under a counterclockwise rotation they pick up a phase:
\be
e^{(\pm)}_i  \mapsto e^{\mp i\theta}e^{(\pm)}_i.
\ee
These are states of definite helicity, the usual circular polarizations in electromagnetism. Note that they are null but not orthogonal
\be
e^{(\pm)}\cdot e^{(\pm)}  = 0 ~~~~~~~~~e^{(\pm)}\cdot e^{(\mp)} = 1.
\ee
To create helicity $\pm J$ states, we take a product of $J$ of the $e^{(\pm)}_i$,
\be
e^{(\pm J)}_{i_1\cdots i_J} = e^{(\pm )}_{i_1}\cdots e^{(\pm )}_{i_J}\,.
\ee
These tensors are traceless and symmetric by virtue of the fact that the $e^{(\pm)}$ are null. These tensors are not orthogonal, but if we take the linear combinations
\begin{align}
e^{(\oplus_J)}_{i_1\cdots i_J} &= \frac{1}{\sqrt 2}\left(e^{(+J)}_{i_1\cdots i_J}+e^{(-J)}_{i_1\cdots i_J}\right)\, ,\\
e^{(\otimes_J)}_{i_1\cdots i_J} &= -\frac{i}{\sqrt 2}\left(e^{(+J)}_{i_1\cdots i_J}-e^{(-J)}_{i_1\cdots i_J}\right),
\end{align}
these are orthonormal,
\be
e^{(\oplus_J)}\cdot e^{(\oplus_J)} = e^{(\otimes_J)}\cdot e^{(\otimes_J)}=1 ~~~~~~~~~e^{(\oplus_J)}\cdot e^{(\otimes_J)} = 0.
\ee
It will be useful to have expressions for contractions of the polarization tensors (when $J_2 > J_1$):
\begin{align}
\label{eq:polcontract1}
e^{(\oplus_{J_1})}_{i_1\cdots i_{J_1}}e^{(\oplus_{J_2})}_{i_1\cdots i_{J_1} j_{J_1+1}\cdots j_{J_2}} &= \frac{1}{\sqrt{2}}e^{(\oplus_{J_2-J_1})}_{ j_{J_1+1} \cdots j_{J_2}}
\, , & e^{(\otimes_{J_1})}_{i_1\cdots i_{J_1}}e^{(\otimes_{J_2})}_{i_1\cdots i_{J_1} j_{J_1+1}\cdots j_{J_2}} &= \frac{1}{\sqrt{2}}e^{(\oplus_{J_2-J_1})}_{ j_{J_1+1} \cdots j_{J_2}}\\
e^{(\oplus_{J_1})}_{i_1\cdots i_{J_1}}e^{(\otimes_{J_2})}_{i_1\cdots i_{J_1} j_{J_1+1}\cdots j_{J_2}} &= \frac{1}{\sqrt 2}e^{(\otimes_{J_2-J_1})}_{j_{J_1+1}\cdots j_{J_2}}\, ,&e^{(\otimes_{J_1})}_{i_1\cdots i_{J_1}}e^{(\oplus_{J_2})}_{i_1\cdots i_{J_1} j_{J_1+1}\cdots j_{J_2}} &= -\frac{1}{\sqrt 2}e^{(\otimes_{J_2-J_1})}_{j_{J_1+1}\cdots j_{J_2}}
\, .
\label{eq:polcontract2}
\end{align}

\section{Eikonal amplitude}
\label{app:eikampcalc}

Here we write out the explicit form of the eikonal amplitude. 
 It is convenient to organize the phase shift~\eqref{eq:curvcubedslJ} as a matrix of the $4^2\times4^2$ possible polarization combinations.

We use the basis for the polarizations described in Appendix~\ref{app:polarizations}. It is also helpful to make the following definition involving derivatives with respect to the impact parameter and the polarization vectors from which the polarization tensors are built:
\be
\vec e^{\,(\pm)}\cdot \vec\partial_b \equiv \frac{1}{\sqrt 2}D^\pm= \frac{1}{\sqrt 2}\left(\partial_{b^1}\pm i\partial_{b^2}\right).
\ee
In simplifying following expressions, it is useful to note that $D^\pm D^\mp  = \nabla^2$ is the two-dimensional Laplacian.

With these definitions, the contraction between the $\oplus$ or $\otimes$ polarization tensors of a spin-$J$ particle and  impact parameter derivatives is given by
\begin{align}
\label{eq:oplusJderiv}
e^{(\oplus_J)}_{i_1\cdots i_J} \partial_{b^{i_1}}\cdots \partial_{b^{i_J}} &= \frac{1}{2^\frac{J+1}{2}}\left[ \left(D^+\right)^J+ \left(D^-\right)^J\right]\\
e^{(\otimes_J)}_{i_1\cdots i_J} \partial_{b^{i_1}}\cdots \partial_{b^{i_J}} &= -\frac{i}{2^\frac{J+1}{2}}\left[ \left(D^+\right)^J- \left(D^-\right)^J\right].
\label{eq:otimesJderiv}
\end{align}
With these simplifications, the amplitude~\eqref{eq:curvcubedslJ} can be written for explicit polarizations as
\be
\delta \simeq \frac{g^2}{4}\frac{s^{J_3-1}}{2^{ J_1+J_2+J_3}M^2\Lambda^{2J_1+2J_2+2J_3-4}} {\bf P}_{3,4}^{\rm T}\,\hat {\cal S}\,{\bf P}_{1,2}\frac{1}{2\pi}\log\left(\frac{L}{b}\right)
\ee
Here ${\bf P}_{a,b}$ is a vector of incoming/outgoing polarization coefficients and $\hat {\cal S} = {\bf S}\otimes {\bf S}$ where ${\bf S}$ is the $4\times 4$ matrix
\scriptsize
\be
{\bf S} = 
\left(
\begin{array}{cccc}
0& 0 & \left(D^+\right)^{J_1+J_2}+ \left(D^-\right)^{J_1+J_2}&i \left(D^-\right)^{J_1+J_2}-i\left(D^+\right)^{J_1+J_2}\\
0& 0& i \left(D^-\right)^{J_1+J_2}-i\left(D^+\right)^{J_1+J_2} &  -\left(D^+\right)^{J_1+J_2}- \left(D^-\right)^{J_1+J_2}\\
\left(D^+\right)^{J_1+J_2}+ \left(D^-\right)^{J_1+J_2}&i \left(D^-\right)^{J_1+J_2}-i\left(D^+\right)^{J_1+J_2} & 0 & 0\\
i \left(D^-\right)^{J_1+J_2}-i\left(D^+\right)^{J_1+J_2} &  -\left(D^+\right)^{J_1+J_2}- \left(D^-\right)^{J_1+J_2} & 0 & 0
\end{array}
\right).
\ee
\normalsize
Using the formula 
\be
\label{eq:derivsonlog}
\left(D^\pm\right)^{\alpha} \log\left(\frac{L}{b}\right) = (-1)^\alpha (2\alpha-2)!! \frac{1}{(b^1\mp ib^2)^\alpha}
\ee
 and choosing the impact parameter to lie along the $1$ direction yields~\eqref{eq:leadingeigenvalues} after diagonalization.

Once we have set $g=0$. the leading amplitude is given by~\eqref{eq:finaleikamp}. In the case where $J_2 = J_1$ the amplitude is diagonal and the eikonal phase shift is manifestly positive. In the other case where $J_2 > J_1$, we can use the formulae~\eqref{eq:polcontract1}--\eqref{eq:polcontract2}, in addition to the ones used at leading order, we obtain the following expression for the eikonal amplitude in terms of explicit polarizations
\be
\delta  = \frac{s^{J_3-1}}{2^{J_2-J_1+2}M^2 \Lambda^{2J_2+2J_3-2J_1-4}}{\bf P}_{3,4}^{\rm T}\,\hat {\cal S}'\,{\bf P}_{1,2}\frac{1}{2\pi}\log\left(\frac{L}{b}\right)
\ee
where in this case $\hat {\cal S}' = \tilde{\bf S}\otimes \tilde{\bf S}$ where $\tilde{\bf S}$ is the $4\times 4$ matrix
\scriptsize
\be
\tilde{\bf S} = 
\left(
\begin{array}{cccc}
0& 0 & \left(D^+\right)^{J_2-J_1}+ \left(D^-\right)^{J_2-J_1}&i \left(D^-\right)^{J_2-J_1}-i\left(D^+\right)^{J_2-J_1}\\
0& 0& i\left(D^+\right)^{J_2-J_1}-i \left(D^-\right)^{J_2-J_1} &  \left(D^+\right)^{J_2-J_1}+ \left(D^-\right)^{J_2-J_1}\\
\left(D^+\right)^{J_2-J_1}+ \left(D^-\right)^{J_2-J_1}&i\left(D^+\right)^{J_2-J_1}-i \left(D^-\right)^{J_2-J_1} & 0 & 0\\
i \left(D^-\right)^{J_2-J_1}-i\left(D^+\right)^{J_2-J_1} &  \left(D^+\right)^{J_2-J_1}+ \left(D^-\right)^{J_2-J_1} & 0 & 0
\end{array}
\right).
\ee
\normalsize
Using equation~\eqref{eq:derivsonlog} again, we obtain the phase shifts~\eqref{eq:subleadingphase}.

\renewcommand{\em}{}
\bibliographystyle{utphys}
\addcontentsline{toc}{section}{References}
\bibliography{masslesseikonal9}

\end{document}